\documentclass{article} 
\usepackage{iclr2027_conference,times}

\usepackage{amsmath,amsfonts,bm}

\def\eqref#1{equation~\ref{#1}}

\def\1{\bm{1}}

\DeclareMathAlphabet{\mathsfit}{\encodingdefault}{\sfdefault}{m}{sl}
\SetMathAlphabet{\mathsfit}{bold}{\encodingdefault}{\sfdefault}{bx}{n}

\usepackage{hyperref}
\usepackage{url}
\usepackage{booktabs}
\usepackage{graphicx} 
\usepackage{wrapfig}
\usepackage{natbib}
\usepackage{subcaption}
\usepackage{multirow}
\usepackage{array}
\usepackage{siunitx}
\usepackage{booktabs}
\usepackage{multirow}
\usepackage{siunitx}
\usepackage{threeparttable}
\usepackage[table]{xcolor}
\usepackage{adjustbox}

\title{MARC: Multi-Bit Watermarking for Autoregressive Audio Generation against Codec Attacks}
   
\author{Liaoran Xu, Weizhi Liu, Zhaoxia Yin \\
East China Normal University, Shanghai, China\\
}

\iclrfinalcopy 
\begin{document}

\maketitle
\fancyhead{}
\renewcommand{\headrulewidth}{0pt}

\begin{abstract}
Generated audio is now used in a range of applications, creating a need to verify its origin after distribution and signal processing. This task is particularly challenging for autoregressive audio generation because codec processing can alter the token sequence recovered from the waveform. Such changes reduce the reliability of watermark detection and payload decoding. Existing methods construct token groups using either intrinsic token representations or substitution patterns caused by transformations. As a result, intrinsic token relationships and codec induced substitutions are modeled separately. In addition, most methods support only zero bit detection. They can determine whether a watermark is present but cannot distinguish individual generated outputs. We propose \textbf{MARC}, a multi-bit generative watermarking method for autoregressive audio generation. MARC integrates intrinsic token representations with confusion patterns obtained through retokenization and multiple codecs, forming a codec-aware token-cluster space. Within this space, payload-driven cluster scheduling is used to embed a multi-bit watermark, while detection and payload decoding are performed on retokenized observations. Experiments on speech, dialogue, and music generation show that MARC achieves an average of 97.3\% bit extraction accuracy on unmodified watermarked audio and the watermark can still be extracted under diverse codec attacks. MARC also demonstrates robustness to overwriting attacks.

\end{abstract}

\section{Introduction}
Autoregressive models are widely used for audio generation \citep{rouard2026continuous}, including speech synthesis \citep{nishimura2025hall}, spoken dialogue \citep{tan2026drvoice}, and music generation \citep{pmlr-v267-liu25m}, where discrete audio representations allow different generation tasks to be formulated as token-sequence modeling. As these models are increasingly used in practical applications, generated audio needs not only reliable source verification but also the ability to distinguish and trace individual outputs \citep{pmlr-v235-san-roman24a}. Multi-bit generative watermarking provides this capability by embedding additional information into each generated audio sample beyond zero-bit detection. This information should remain recoverable after the audio is released. In reality, generated audio may undergo compression or other codec processing during transmission, storage, and redistribution \citep{NEURIPS2024_5d9b7775}. Verification for autoregressive token-level watermarking requires retokenizing the generated waveform, and the resulting token sequence may differ from the one used during generation. Codec processing can further increase these token substitutions, making both watermark detection and multi-bit recovery more difficult.

\begin{figure}[h]
  \centering
  \includegraphics[width=\linewidth]{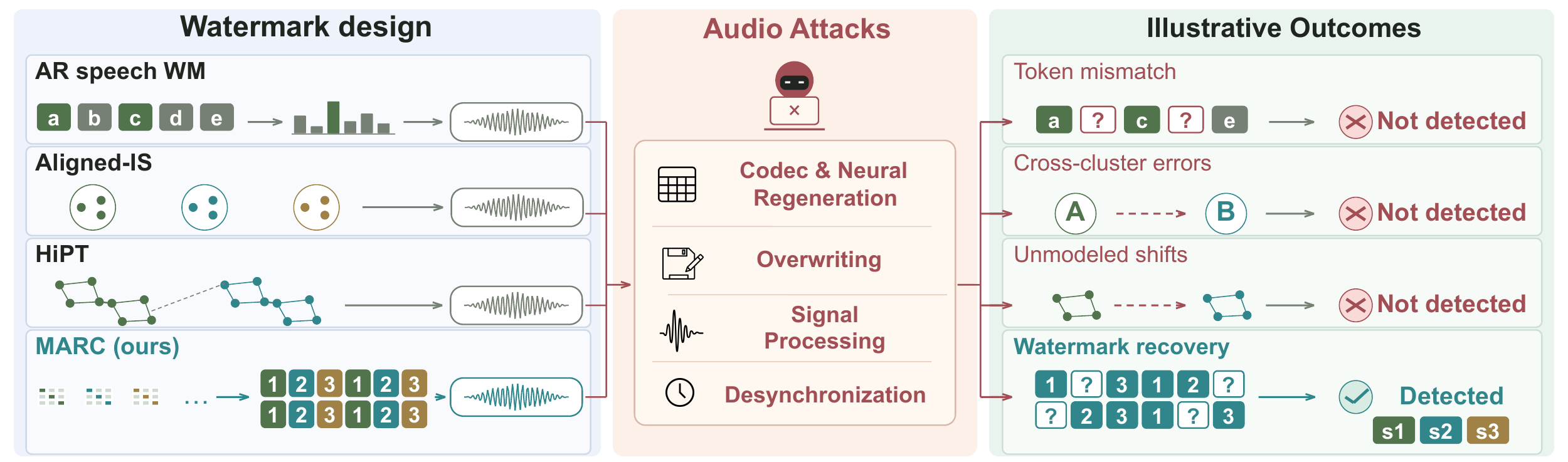}
  \caption{Method overview.}
  \label{abstract}
\end{figure}

Existing work has developed several forms of generative watermarking for autoregressive and token-based audio generation as shown in Figure \ref{abstract}. \citet{wu25k_interspeech} introduce statistical token-level watermarking and identifies retokenization mismatch. Aligned-IS \citep{NEURIPS2025_9f5ca3d3} and Hidden in Plain Tokens (HiPT) \citep{milis2026hidden} further improve robustness through token grouping, but two limitations remain. First, \textbf{single-source token grouping}: existing clustering strategies rely on a single source of token relationships: Aligned-IS groups tokens using intrinsic representations, whereas HiPT derives groups from transformation-induced confusion, so these two types of information are not modeled jointly. Second, \textbf{zero-bit limitation}: existing autoregressive watermarking primarily targets zero-bit detection, which can detect watermark but cannot carry additional information for distinguishing individual generated outputs. Extending such methods to multi-bit watermarking is difficult because retokenization and codec processing can directly corrupt the embedded bits.
To solve these problems, we propose \textbf{MARC}, a generative watermarking framework that enables robust multi-bit embedding for autoregressive audio generation. To address \textbf{single-source token grouping}, MARC combines intrinsic token representations with confusion patterns induced by retokenization and diverse codecs to build a codec-aware token-cluster space. To address \textbf{zero-bit limitation}, MARC uses payload-driven cluster scheduling to embed a multi-bit watermark and performs detection and decoding on retokenized observations to recover both watermark presence and the embedded bits. 
MARC additionally introduces a public check value (pub-key) and a secret auxiliary suffix (pri-key) for message consistency checking, with both parameters used in the full verification and recovery protocol without assuming cryptographic security. 
Across multiple autoregressive audio generation tasks, MARC achieves an average of 97.3\% bit extraction accuracy on unmodified watermarked audio and the watermark can still be extracted under diverse codec attacks.
Our contributions are summarized as follows:
\begin{itemize}
    \item We propose \textbf{MARC}, a codec-robust multi-bit generative watermarking framework for autoregressive audio generation that extends zero-bit detection to multi-bit payload embedding and recovery.

    \item We propose a \textbf{codec-aware token-cluster representation} that combines intrinsic token structure with codec-aware confusion, reducing the limitations of single-source token grouping under different codec transformations.

    \item We develop \textbf{payload-driven embedding and robust recovery} for multi-bit watermarking under token-level errors. Experiments on speech, dialogue, and music generation show that MARC works across different autoregressive audio generation tasks and remains robust under codec attacks.

    \item We demonstrate \textbf{robustness to watermark overwriting}, showing that MARC preserves detection and recovery of the original payload when an additional watermark is embedded into the released audio.
\end{itemize}

\section{Related Work}

\paragraph{Generative Audio Watermarking.}
Generative audio watermarking embeds watermark information during audio synthesis rather than adding it through post-hoc waveform modification. Existing methods integrate watermarks into diffusion generation \citep{cao2023invisible,liu2024groot}, latent representations \citep{san2025latent}, diffusion noise distributions \citep{tang2025poisoning}, model parameters \citep{feng2025robust}, or diffusion-domain modulation \citep{lee2025robust}. These studies show that watermarking can be incorporated into different audio generators, but their mechanisms mainly depend on continuous latent, diffusion, or model-specific representations and are not directly applicable to discrete token sampling in autoregressive generation. Autoregressive audio watermarking instead operates on token distributions and must remain detectable after the generated waveform is retokenized.

\paragraph{Autoregressive Audio Watermarking.}
TraceableSpeech \citep{zhou24b_interspeech} generates watermarked speech within TTS, while \citet{wu25k_interspeech} introduces statistical token-level watermarking for autoregressive speech and identifies retokenization mismatch as a central issue. Aligned-IS \citep{NEURIPS2025_9f5ca3d3} reduces this mismatch by clustering tokens according to intrinsic representation geometry. However, its clustering does not explicitly consider codec attacks. HiPT \citep{milis2026hidden} instead builds token groups from transformation-aware confusion, directly modeling token substitutions observed after processing, but does not jointly use intrinsic token structure. The two methods therefore rely on different sources of token relationships, but they do not model these sources jointly. In addition, existing autoregressive methods mainly focus on zero-bit detection rather than embedding multi-bit information. Joint modeling of token structure and codec-aware confusion with robust multi-bit recovery still remains insufficiently addressed. Our method combines these two sources of token information and extends autoregressive generative watermarking to multi-bit embedding and recovery.

\section{Threat Model}

\paragraph{Adversary Goal.}
We consider a waveform $y_w$ generated by an autoregressive
audio model and watermarked with a multi-bit payload $\mathbf{m}$. The full verification protocol additionally uses a public check value (pub-key) and a secret auxiliary suffix
(pri-key) for message consistency checking, and neither
parameter is assumed to provide cryptographic meaning. After release, an output-level transformation $\mathcal{A}$
produces
\[
\tilde{y}=\mathcal{A}(y_w),
\]
which is retokenized for watermark verification and payload
recovery.
The verifier uses $\mathbf{m}^{\star}$ to search over candidate
alignments and then estimates the payload from the selected
observations.
The adversary aims to cause missed detection or incorrect
payload recovery while preserving the underlying audio content
and perceptual usability.
The same setting covers benign post-processing. 

\paragraph{Adversary Capability.}
The pri-key is not directly disclosed, only the owner and the authorized user can have it. The adversary does not control the verifier's payload hypothesis $\mathbf{m}^{\star}$.
We consider neural audio codec attacks, signal processing, and desynchronization applied to the released waveform.
These transformations can alter both token identities and their
temporal correspondence with watermark embedding steps, even when
perceptual content is preserved.
MARC focuses on multi-bit verification and recovery under this
joint token mismatch and temporal misalignment, with codec-induced
retokenization mismatch as the primary challenge.
Test-time codecs may include those unseen during construction of
the codec-aware cluster map.

\begin{figure}[h]
  \centering
  \includegraphics[width=\linewidth]{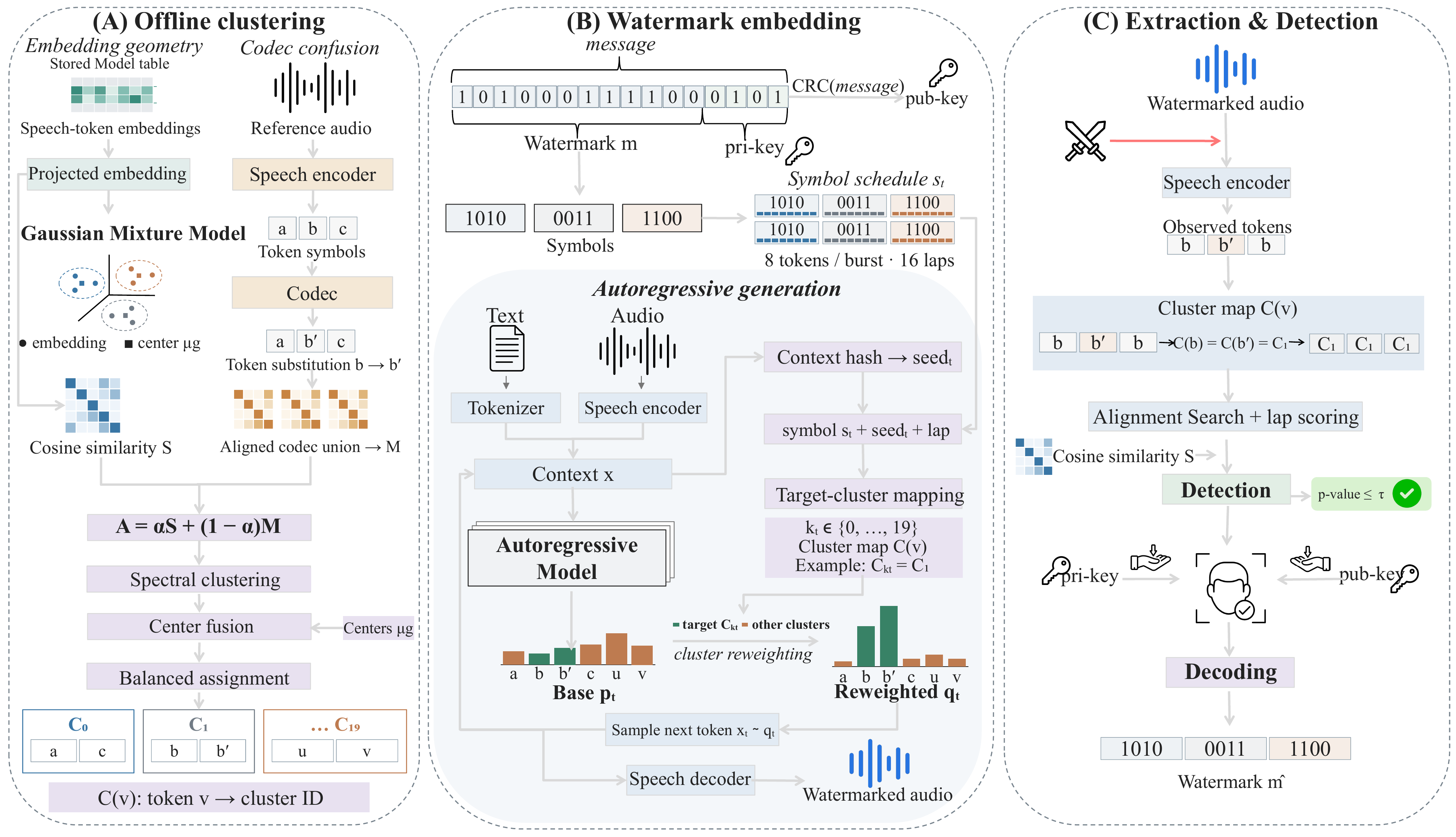}
  \caption{Method pipeline: (A) \textit{Offline clustering}: joint embedding and code similarities yield the token-to-cluster map $C(v)$. (B) \textit{Watermark embedding}: key-conditioned symbol scheduling and cluster reweighting guide autoregressive generation of watermarked audio. (C) \textit{Extraction and detection}: observed tokens are cluster-mapped and alignment-scored for watermark detection and message decoding.}
  \label{pip}
\end{figure}

\section{MARC: Codec-Robust Multi-bit Watermarking}
\label{sec:marc}

\paragraph{Analysis.}

Codec re-encoding changes discrete token IDs and can disrupt temporal alignment. Recent methods encode watermarks through frame-level cluster membership and detect them using $S=\sum_t\mathbf{1}[c(y_t)=s_t]$. Codec-induced mismatches can drive $S$ toward chance level. Zero-bit methods can pool $s_t$ into a global set for position-independent counting, whereas multi-bit recovery requires alignment with embedding positions: an alignment error can propagate to subsequent bits, while finer clustering increases vulnerability to codec-induced boundary crossings.

MARC merges tokens that would swap places under the codec into the same cluster \(c_\cup\), allowing the symbols themselves to survive compression, and then uses local shifts to realign misaligned frames $\hat{\delta}=\arg\max_{\delta}\sum_t \mathbf{1}[c_{\cup}(y_{t+\delta})=s_t(m)].$ 
MARC combines complementary token relationships to support multi-bit watermarking under retokenization and codec processing.
Offline clustering fuses embedding geometry with empirical codec
confusion to construct a codec-aware cluster map.
During generation, a repeated symbol schedule selects keyed target
clusters for probability reweighting. Users can detect whether the audio contains a watermark. The legal verifier uses both pub-key and pri-key to recover watermark. Our method pipeline is shown in Figure \ref{pip}.

\subsection{Codec-Aware Clustering}

\label{sec:marc-clustering}

\paragraph{Embedding Geometry.}
We project the generator's audio-token embeddings into a reduced
representation space and fit a Gaussian mixture model to obtain
geometric centers $\{\mu_k^{\mathrm G}\}_{k=0}^{K-1}$.
For projected embeddings $e_a$ and $e_b$, we define
$S_{ab}=(1+\cos(e_a,e_b))/2$.
This representation provides token relationships independently of
which transformations appear in the calibration data.

\paragraph{Codec Confusion.}
Let $E$ and $D$ denote the speech encoder and decoder.
For each reference waveform $x\in\mathcal D$, we obtain
$u=E(x)$, the reconstruction sequence $v^{(0)}=E(D(u))$,
and codec-processed sequences $v^{(j)}=E(\mathcal A_j(x))$.
We accumulate confusion counts \citep{mangu2000finding} by pairing tokens at matching
indices over the common sequence length:
\begin{equation}
M_{ab}^{(j)}
=
\sum_{x\in\mathcal D}
\sum_{i=1}^{\min(|u|,|v^{(j)}|)}
\mathbf 1[u_i=a,\ v_i^{(j)}=b].
\label{eq:confusion_count}
\end{equation}
These counts capture empirical changes across the reconstruction
and codec channels, including residual temporal mismatch.

We symmetrize and normalize each matrix using
$\mathcal N(M)=(M+M^\top)/\max_{a,b}(M+M^\top)_{ab}$,
leaving zero matrices unchanged.
The codec union is
\begin{equation}
M_{\mathrm U}
=
\beta\mathcal N(M^{(0)})
+
(1-\beta)
\max_{0\le j\le J}\mathcal N(M^{(j)}).
\label{eq:codec_union}
\end{equation}
The element-wise maximum retains strong channel-specific
confusions, while $\beta$ adds weight to ordinary retokenization.

\paragraph{Fusion and Cluster Assignment.}
We combine embedding similarity and normalized codec confusion,
then apply spectral clustering to the resulting affinity:
\begin{equation}
A=\alpha S+(1-\alpha)\mathcal N(M_{\mathrm U}),
\qquad
\mu_k=\alpha\mu_k^{\mathrm G}
+(1-\alpha)\mu_k^{\mathrm S},
\label{eq:clustering_fusion}
\end{equation}
where $\mu_k^{\mathrm S}$ is the centroid of spectral cluster $k$.
The implementation blends the two center sets in their returned
index order, maps the centers back to the embedding space, and
applies balanced assignment.
This yields a codec-aware cluster map
$C:\mathcal V_a\rightarrow\{0,\ldots,K-1\}$, hereafter the
cluster map.
We use $K=20$.
Balanced assignment controls cluster cardinalities rather than
their probability mass under the generator.

\subsection{Watermark Embedding}
\label{sec:marc-embedding}

\paragraph{Message and Symbol Schedule.}
The embedded payload $\mathbf m\in\{0,1\}^{12}$ is partitioned
into three symbols $s_0,s_1,s_2\in\{0,\ldots,15\}$.
The main configuration assigns 8 consecutive watermark steps
to each symbol, cycles through the three positions, and repeats
the schedule across 16 laps.
The public check value is computed as \(k_{\rm pub}=\mathrm{CRC}(m,k_{\rm pri})\), where \(k_{\rm pri}\) denotes the secret auxiliary suffix shared by the owner and the authorized verifier.

\paragraph{Target-cluster Mapping.}
Text or audio prompts form the context of the autoregressive
generator. At watermark step $t$, the secret key $\kappa$ and preceding token produce a context seed $r_t$.
Let $j_t$ and $\ell_t$ denote the scheduled symbol position and
lap index.
The target cluster is
\begin{equation}
k_t
=
\bigl[
b(s_{j_t})+
H(s_{j_t},r_t,\ell_t,j_t)
\bigr]\bmod K,
\label{eq:payload_cluster}
\end{equation}
where $b$ is a fixed symbol-to-cluster base mapping and $H$ is
a domain-separated hash.
The mapping omits the exact within-burst step index.
Changes in context or lap can nevertheless assign repeated
symbols to different target clusters.

\paragraph{Distribution Reweighting.}
Let $p_t(v)$ be the generator's next-token distribution.
We extend the audio cluster map to the sampling vocabulary by
assigning non-audio tokens to the remaining groups.
For $V_k=\{v:C(v)=k\}$, define
$P_{t,k}=\sum_{v\in V_k}p_t(v)$.
The reweighting rule \citep{pmlr-v235-wu24h} gives
\begin{equation}
q_t(v)
=
p_t(v)\frac{\widehat P_{t,C(v)}}{P_{t,C(v)}},
\qquad
\widehat P_{t,k_t}=\min(KP_{t,k_t},1).
\label{eq:distribution_reweighting}
\end{equation}
When the target mass is below one, the remaining mass is distributed
among non-target groups in proportion to $(KP_{t,k}-1)_+$.
Zero-probability groups contribute no mass.
This changes cluster probabilities while preserving relative
probabilities within each group.
The generator samples from $q_t$, and the speech decoder converts
the generated sequence into watermarked audio.

\subsection{Extraction and Detection}
\label{sec:marc-decoding}

\paragraph{Alignment Search and Lap Scoring.}
The verifier receives the processed waveform $\widetilde{y}$,
the secret key $\kappa$, and the claimed payload $\mathbf m^\star$.
Speech encoding produces observed tokens $\widetilde z_t$, which
are mapped to clusters $\widetilde c_t=C(\widetilde z_t)$.
The verifier also recomputes context seeds $\widetilde r_t$.
Tokens mapped to the same cluster preserve the current cluster
observation even when their identities differ.

The verifier evaluates waveform variants, observation windows,
and offsets of the watermark-step indices.
Waveform variants are retokenized separately.
Index offsets change the symbol and lap assignments while retaining the observed clusters and their associated seeds.
The evaluation can use the supplied attack type and configured
processing parameters to construct waveform variants.

For a window and offset $\delta$, the claimed payload determines
the expected cluster sequence.
We compute
\begin{equation}
\begin{aligned}
h_\delta
&=
\sum_t
\mathbf 1[
\widetilde c_t=
k_{t+\delta}(\mathbf m^\star,\widetilde r_t)],\\
p_\delta
&=
\Pr[
\operatorname{Bin}(N_\delta,1/K)\ge h_\delta],
\end{aligned}
\label{eq:alignment_detection}
\end{equation}
where $N_\delta$ is the number of observations.
Detection retains the strongest candidate across windows and
waveform variants, while decoding evaluates the candidate offsets
individually.

\paragraph{Detection.}
The detector compares the selected score with a threshold $\tau$.
It can augment hard matches with capped soft evidence from an
auxiliary cluster affinity matrix
$R\in\mathbb R^{K\times K}$ and optionally combine the resulting
scores.
Here $R$ scores relations between observed and expected clusters,
whereas $A$ is the token affinity used for offline clustering.

\paragraph{Decoding.}
For each candidate alignment $a$, the decoder enumerates all
16 values at every symbol position.
Let $\mathcal O_a$ denote its observations and let $\eta_a$ be the
match rate against $\mathbf m^\star$, clipped to $[0.05,0.95]$.
The hard symbol score is
\begin{equation}
L_{q,a}^{\mathrm{hard}}(s)
=
\sum_{\substack{t\in\mathcal O_a\\j_t=q}}
\log\!\left[
\eta_a\mathbf 1[
\widetilde c_t=k_t(s,\widetilde r_t)]
+\frac{1-\eta_a}{K}
\right].
\label{eq:symbol_score}
\end{equation}
The optional soft rule gives partial compatibility to unequal
clusters when
$R_{ij}/\max(R_{ii},R_{jj},\epsilon)$ exceeds its configured
threshold.
For each rule, the decoder maximizes the symbol score independently
at each position.
It retains the rule with the larger summed maximum score and
converts the resulting symbols into a candidate bit string
$\widehat{\mathbf m}^{(a)}$.

The pri-key scores candidate payloads before aggregation.
The resulting $n$ selected candidates are combined by bit-wise
majority voting. The voted payload initializes a local refinement
guided by pooled symbol scores and the pub-key:
\begin{equation}
\begin{aligned}
m_b^{(0)}
&=
\mathbf 1\!\left[
\sum_{a=1}^{n}\widehat m_b^{(a)}>n/2
\right],
\\
J(\mathbf s)
&=
\sum_q\overline L_q(s_q)
-\gamma\left[
1-V_{\mathrm{pub}}(\mathcal B(\mathbf s),k_{\mathrm{pub}})
\right].
\end{aligned}
\label{eq:vote_crc}
\end{equation}
Here, $\mathcal B$ maps symbols to a payload, and $V_{\mathrm{pub}}$
returns one if the payload passes pub-key verification and
zero otherwise. The pooled score $\overline L_q$ aggregates
symbol evidence across alignments, and $\gamma>0$ penalizes
failed verification. Voting ties resolve to zero.
Local search evaluates high-scoring symbol replacements and
accepts the refined payload $\widehat{\mathbf m}$ only if
it passes pub-key verification.

\section{Evaluation}
\label{sec:evaluation}

\subsection{Experimental Setup}
\label{setup}
For our main audio generation experiments, we consider two audio generative models with autoregressive architecture, that span different applications and use different audio codecs for discretization. For dialogue and speech synthesis tasks, We evaluate MARC on SpiritLM \citep{nguyen-etal-2025-spirit} using Finance QA \citep{tu2024waterbench}, Dolly CW \citep{conover2023free}, and LibriSpeech \citep{panayotov2015librispeech}. For music generation tasks we evaluate on MusicGen \citep{copet2023simple} using LibriSpeech, MusicCaps \citep{agostinelli2023musiclm}, and MTG-Jamendo \citep{bogdanov2019mtg}.

We compare our method with KGW using $\delta=2.0$ following \citet{wu25k_interspeech} which we name as AR speech WM, Aligned-IS \citep{NEURIPS2025_9f5ca3d3}, and HiPT \citep{milis2026hidden}.
As for our method MARC, embeds a 12 bit payload by default. The main configuration assigns 8 consecutive watermark steps
to each symbol, cycles through the three positions, and repeats
the schedule across 16 laps. 

We evaluate multi-bit watermark detection using bit accuracy as the detection score, True Positive Rate (TPR) is measured at a target False Positive Rate (FPR) of $1\%$. Cluster Median-p (cluster\_p) summarizes the cluster-based detection statistic. Bit accuracy (bit\_acc) is the fraction of correctly recovered payload bits.
For fidelity, we report predicted MOS with NISQA \citep{mittag21_interspeech} and DNSMOS \citep{cumlin2024dnsmos} scores for speech outputs. We compute FAD using VGGish features \citep{Kilgour2019FrchetAD}. NISQA and DNSMOS are not used to assess generated music. These two scores correspond to the naturalness of speech and noise levels, which do not measure musical quality, and therefore MusicGen’s results do not use these two metrics.

\definecolor{marcrow}{gray}{0.94}

\begin{table*}[t]
\centering
\caption{
Watermark reliability and audio fidelity on SpiritLM and MusicGen:
All methods except MARC are zero-bit watermarking methods,
so bit-level TPR and bit accuracy are not applicable to these baselines.
NISQA and DNSMOS assess speech quality and are not applicable
to music generation.
}
\label{tab:main-results}

\begingroup
\small
\setlength{\tabcolsep}{4pt}
\renewcommand{\arraystretch}{1.05}
\sisetup{
  detect-weight = true,
  detect-family = true
}

\begin{adjustbox}{max width=\linewidth}
\begin{tabular}{
  @{}ll
  S[table-format=1.2]
  S[table-format=2.2]
  S[table-format=1.2]
  S[table-format=1.3]
  S[table-format=1.3]
  S[table-format=1.3]
  @{}
}
\toprule
\multirow{2}{*}{Dataset / Task}
& \multirow{2}{*}{Method}
& \multicolumn{3}{c}{Reliability}
& \multicolumn{3}{c}{Fidelity} \\
\cmidrule(lr){3-5}
\cmidrule(lr){6-8}
&
& \multicolumn{1}{c}{TPR $\uparrow$}
& \multicolumn{1}{c}{-log($cluster\_p$) $\uparrow$}
& \multicolumn{1}{c}{bit\_acc $\uparrow$}
& \multicolumn{1}{c}{FAD $\downarrow$}
& \multicolumn{1}{c}{NISQA $\uparrow$}
& \multicolumn{1}{c}{DNSMOS $\uparrow$} \\
\midrule

\multicolumn{8}{c}{\textbf{(a) SpiritLM}} \\

\multirow{4}{*}{%
  \shortstack[l]{Finance QA\\{\footnotesize (Dialogue)}}}
& AR speech WM
& {--} & 3.54 & {--}
& 0.075 & 4.505 & 3.854 \\
& Aligned-IS
& {--} & 4.47 & {--}
& \bfseries 0.056 & \bfseries 4.519 & 3.845 \\
& HiPT
& {--} & 21.44 & {--}
& 1.031 & 4.429 & 3.772 \\
\rowcolor{marcrow}
\cellcolor{white}
& \textbf{MARC (Ours)}
& 0.94 & \bfseries 37.08 & 0.98
& 0.262 & 4.503 & \bfseries 3.868 \\
\midrule

\multirow{4}{*}{%
  \shortstack[l]{Dolly CW\\{\footnotesize (Dialogue)}}}
& AR speech WM
& {--} & 3.31 & {--}
& 0.077 & 4.491 & 3.831 \\
& Aligned-IS
& {--} & 5.33 & {--}
& 0.055 & 4.514 & 3.791 \\
& HiPT
& {--} & 17.81 & {--}
& 0.573 & \bfseries 4.516 & 3.794 \\
\rowcolor{marcrow}
\cellcolor{white}
& \textbf{MARC (Ours)}
& 0.97 & \bfseries 34.16 & 0.99
& \bfseries 0.024 & 4.491 & \bfseries 3.867 \\
\midrule

\multirow{4}{*}{%
  \shortstack[l]{LibriSpeech\\{\footnotesize (Speech generation)}}}
& AR speech WM
& {--} & 3.22 & {--}
& 0.063 & \bfseries 4.501 & 3.840 \\
& Aligned-IS
& {--} & 5.11 & {--}
& \bfseries 0.048 & 4.486 & \bfseries 3.875 \\
& HiPT
& {--} & 9.61 & {--}
& 0.625 & 4.401 & 3.842 \\
\rowcolor{marcrow}
\cellcolor{white}
& \textbf{MARC (Ours)}
& 0.96 & \bfseries 30.01 & 0.98
& 0.123 & 4.446 & 3.867 \\

\specialrule{1.1pt}{3pt}{3pt}

\multicolumn{8}{c}{\textbf{(b) MusicGen}} \\

\multirow{4}{*}{%
  \shortstack[l]{LibriSpeech\\{\footnotesize (Music generation)}}}
& AR speech WM
& {--} & 2.34 & {--}
& 1.062 & {--} & {--} \\
& Aligned-IS
& {--} & 3.58 & {--}
& \bfseries 0.495 & {--} & {--} \\
& HiPT
& {--} & 28.55 & {--}
& 1.171 & {--} & {--} \\
\rowcolor{marcrow}
\cellcolor{white}
& \textbf{MARC (Ours)}
& 0.96 & \bfseries 34.04 & 0.97
& 1.176 & {--} & {--} \\
\midrule

\multirow{4}{*}{%
  \shortstack[l]{MusicCaps\\{\footnotesize (Music generation)}}}
& AR speech WM
& {--} & 1.86 & {--}
& 0.924 & {--} & {--} \\
& Aligned-IS
& {--} & 3.47 & {--}
& \bfseries 0.643 & {--} & {--} \\
& HiPT
& {--} & 25.14 & {--}
& 1.056 & {--} & {--} \\
\rowcolor{marcrow}
\cellcolor{white}
& \textbf{MARC (Ours)}
& 0.95 & \bfseries 26.15 & 0.97
& 1.182 & {--} & {--} \\
\midrule

\multirow{4}{*}{%
  \shortstack[l]{MTG-Jamendo\\{\footnotesize (Music generation)}}}
& AR speech WM
& {--} & 0.47 & {--}
& 1.018 & {--} & {--} \\
& Aligned-IS
& {--} & 3.47 & {--}
& \bfseries 0.596 & {--} & {--} \\
& HiPT
& {--} & 17.04 & {--}
& 1.131 & {--} & {--} \\
\rowcolor{marcrow}
\cellcolor{white}
& \textbf{MARC (Ours)}
& 0.90 & \bfseries 22.90 & 0.95
& 1.616 & {--} & {--} \\
\bottomrule
\end{tabular}
\end{adjustbox}

\endgroup
\end{table*}

\subsection{Reliability and Fidelity}

Table~\ref{tab:main-results} reports detection and payload recovery for each model and dataset pair. The results for SpiritLM are presented in Table \ref{tab:main-results}(a), covering Finance QA, Dolly CW, and LibriSpeech. Across all three datasets, the data shows that our method, MARC, not only outperforms existing approaches in terms of clustering p-values, but also makes the watermark easier to detect, whilst achieving high detection and extraction rates for multi-bit watermarks. As shown by the fidelity data in the table, although our approach involves some trade-offs in terms of fidelity in order to achieve multi-bit representation, it still maintains good audio quality and offers advantages over existing approaches in dialogue tasks, speech synthesis and music generation.

\begin{figure}[t]
  \centering
  \includegraphics[width=0.7\linewidth]{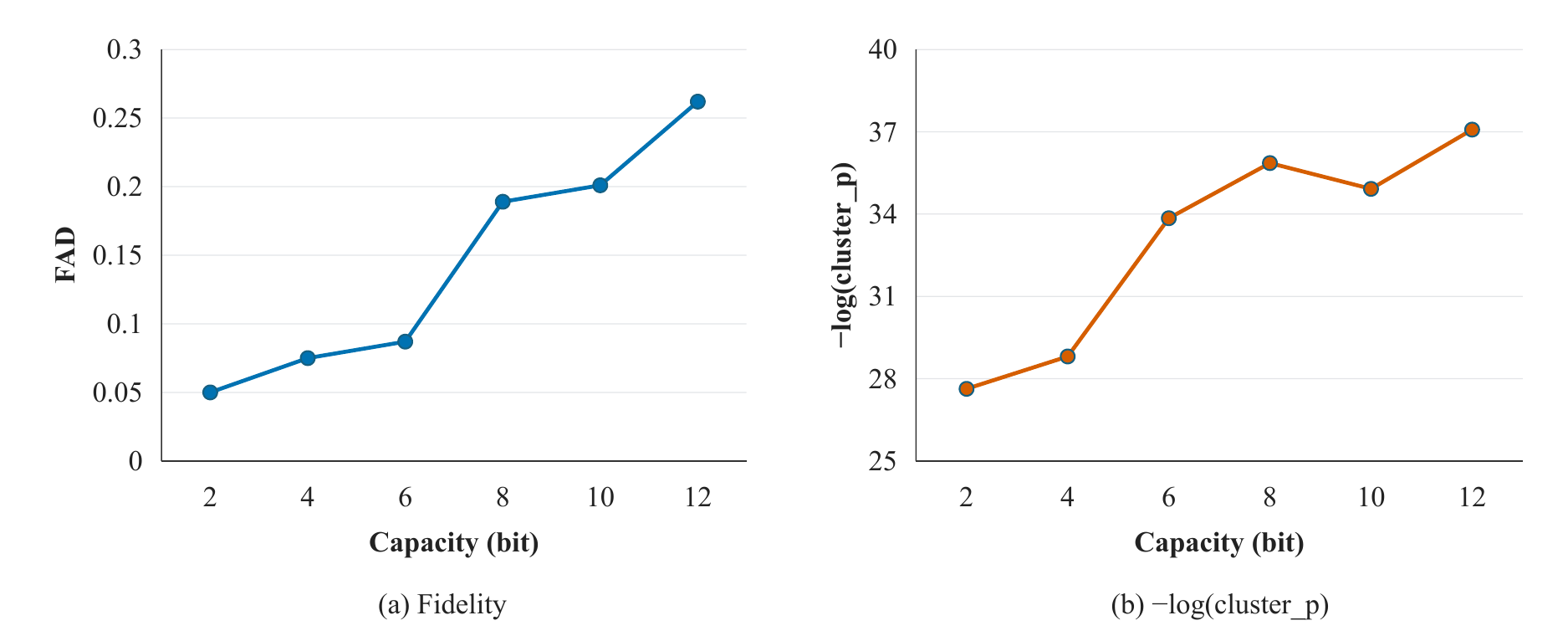}
  \caption{Fidelity and reliability across different capacities with SpiritLM on FinanceQA.}
  \label{Capacities}
\end{figure}

\subsection{Multi-bit Capacity}

Given a fixed audio duration, a watermark token budget and an embedding strength, the payload is divided into several symbol blocks, we use 4-bit symbols when the payload length is divisible by four and 2-bit symbols otherwise. As the number of blocks increases with capacity, the number of observations allocated to each symbol within the same generation decreases.As shown in Figure \ref{Capacities}, FAD increases slightly, as the higher capacity results in denser symbol placement and tighter constraints on clusters imposed by 4-bit symbols. At the same time, as capacity increases, the detection score increases with a decrease at 10 bits. We have opted for a 12-bit embedding here because, following robustness experiments, we found that 12 bits is sufficient to carry the watermark information without spreading the observations per symbol too thinly, thereby maintaining good robustness.

\subsection{Robustness}
We evaluate MARC against two types of attacks: audio transformation attacks, which perturb the watermarked audio, and overwriting attacks, which introduce a competing watermark to challenge the original attribution. Experiments are conducted on SpiritLM and MusicGen. We consider 21 attacks spanning 9 codec and neural regeneration attacks, 8 signal processing attacks, and 4 desynchronization attacks. These transformations challenge token consistency, acoustic evidence preservation, and temporal alignment, respectively. As shown in Table~\ref{tab:robustness}, MARC outperforms all baselines on this metric under all nine codec and neural regeneration attacks on both models. 

\begin{figure}[t]
  \centering
  \includegraphics[width=\linewidth]{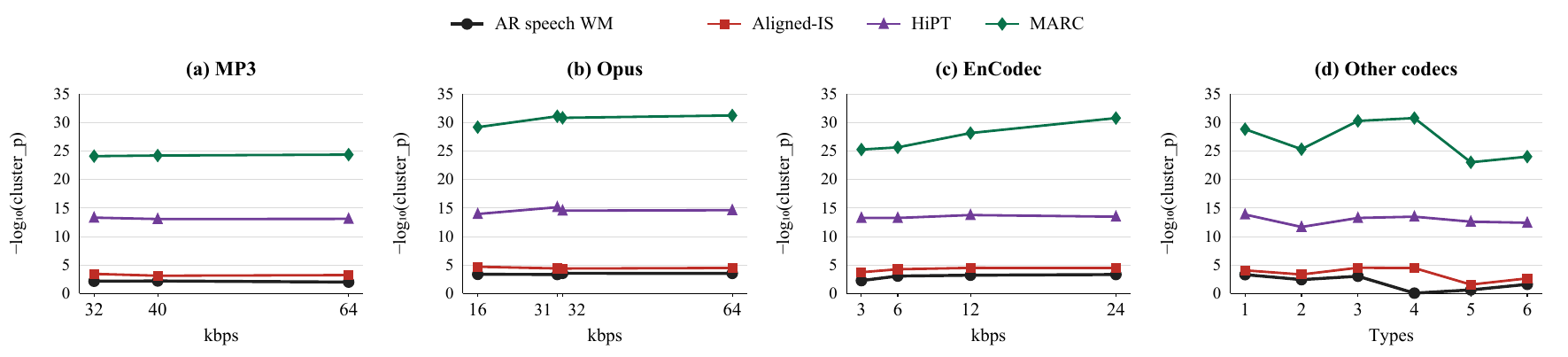}
  \caption{Codec attacks under different parameter: Other codecs: 1 = DAC; 2 = SpeechTokenizer; 3 = FaCodec; 4 = VoiceCraft-EnCodec; 5 = SoCodec; 6 = SemantiCodec.}
  \label{different_parameter}
\end{figure}

This performance is consistent with the complementary roles of payload-driven scheduling and alignment-aware decoding, repeated observations preserve recovery evidence, while alignment correction and cross-lap aggregation accommodate temporal disruption. Additional codec parameter evaluations and experimental details are provided in Figure \ref{different_parameter} and Appendices \ref{Robustness Experiments}.

For the overwriting attack, an overwriting attacker embeds a competing message to replace the original attribution. Table~\ref{tab:fixed_message_replacement} shows that the legitimate owner can still recover the original payload using their private key, and the attacker's message is also recoverable with the attacker's key. This preservation of the original payload demonstrates MARC's resistance to overwriting under the evaluated conditions and supports key-conditioned payload recovery.

\begin{table}[t]
\centering
\caption{Overwriting Attacks: The ‘watermark’ indicator refers to the owner of the original watermark. ‘Verified key’ refers to the type of pri-key held by the person performing the decoding.} 
\label{tab:fixed_message_replacement}
\renewcommand{\arraystretch}{0.8}
\begin{tabular}{lccccc}
\toprule
Attack condition &watermark& Verified key
& TPR $\uparrow$
& $-\log(\mathrm{cluster\_p})$ $\uparrow$
& $\mathrm{bit\_acc}$ $\uparrow$ \\
\midrule
No attack &Owner& Owner
& 0.94 & 37.08 & 0.98 \\
After Overwriting &Owner& Owner
& 0.90 & 24.18 & 0.88 \\
After Overwriting &Owner& Attacker
& 0.68 & 14.18 & 0.75 \\
After Overwriting &Attacker& Attacker
& 0.93 & 35.92 & 0.98 \\
\bottomrule
\end{tabular}
\end{table}

\begin{table*}[t]
\centering
\captionsetup{skip=2pt}
\caption{Robustness under different attacks.}
\label{tab:robustness}
\setlength{\tabcolsep}{3pt}
\renewcommand{\arraystretch}{0.8}

\begin{subtable}{\textwidth}
\centering
\caption{SpiritLM with FinanceQA}
\label{tab:robustness-spiritlm}
\resizebox{\linewidth}{!}{%
\begin{tabular}{l*{6}{c}}
\toprule
\multirow[c]{2}{*}{Attacks}
& AR speech WM
& Aligned-IS
& HiPT
& \multicolumn{3}{c}{MARC(ours)} \\
\cmidrule(lr){2-2}
\cmidrule(lr){3-3}
\cmidrule(lr){4-4}
\cmidrule(lr){5-7}
& $-\log(\mathrm{cluster\_p})$ $\uparrow$
& $-\log(\mathrm{cluster\_p})$ $\uparrow$
& $-\log(\mathrm{cluster\_p})$ $\uparrow$
& TPR $\uparrow$
& $-\log(\mathrm{cluster\_p})$ $\uparrow$
& bit\_acc $\uparrow$ \\
\midrule
\multicolumn{7}{l}{\small\textbf{Codec \& Regeneration}} \\
MP3 (64kbps)       & 2.00 & 3.22 & 13.09 & 0.84 & 24.37 & 0.97 \\
Opus (16kbps)      & 3.36 & 4.39 & 13.95 & 0.96 & 29.20 & 0.99 \\
EnCodec (24kbps)    & 3.32 & 4.36 & 13.48 & 0.97 & 30.79 & 0.99 \\
DAC                & 3.30 & 4.05 & 13.86 & 0.94 & 28.84 & 0.98 \\
SpeechTokenizer    & 2.42 & 3.33 & 11.67 & 0.90 & 25.33 & 0.98 \\
FaCodec            & 3.02 & 4.41 & 13.26 & 0.92 & 30.29 & 0.98 \\
VoiceCraft-EnCodec & 1.02 & 4.46 & 13.47 & 0.97 & 30.70 & 0.95 \\
SoCodec            & 0.62 & 1.56 & 12.60 & 0.78 & 23.03 & 0.96 \\
SemantiCodec       & 1.60 & 2.64 & 12.41 & 0.85 & 24.02 & 0.97 \\
\midrule
\multicolumn{7}{l}{\small\textbf{Signal Processing}} \\
high-pass          & 3.01 & 3.87 & 15.03 & 0.85 & 23.59 & 0.95 \\ 
low-pass           & 3.06 & 3.88 & 14.60 & 0.90 & 28.38 & 0.98 \\
smooth             & 3.42 & 4.54 & 14.47 & 0.94 & 30.80 & 0.99 \\
Gaussian\_noise    & 3.33 & 4.10 & 14.52 & 0.93 & 26.08 & 0.98 \\
pink\_noise        & 3.51 & 4.41 & 14.38 & 0.94 & 30.76 & 0.99 \\
echo               & 3.54 & 4.38 & 14.63 & 0.94 & 28.01 & 0.99 \\
quantization       & 2.45 & 3.52 & 12.48 & 0.91 & 22.86 & 0.98 \\
updownresample     & 3.49 & 4.44 & 14.82 & 0.96 & 31.35 & 0.99 \\
\midrule
\multicolumn{7}{l}{\small\textbf{Desynchronization}} \\
speed              & 0.52 & 1.42 &  6.45 & 0.83 & 18.53 & 0.93 \\ 
crop               & 1.65 & 3.54 &  6.80 & 0.96 & 18.69 & 0.99 \\
shift (1/8)        & 3.09 & 4.14 & 14.83 & 0.91 & 28.31 & 0.98 \\
shush              & 3.46 & 4.37 & 13.82 & 0.95 & 27.14 & 0.99 \\
\bottomrule
\end{tabular}%
}
\end{subtable}

\par\vspace{2pt}

\begin{subtable}{\textwidth}
\centering
\caption{MusicGen with MusicCaps}
\label{tab:robustness-musicgen}
\resizebox{\linewidth}{!}{%
\begin{tabular}{l*{6}{c}}
\toprule
\multirow[c]{2}{*}{Attack}
& AR speech WM
& Aligned-IS
& HiPT
& \multicolumn{3}{c}{MARC(ours)} \\
\cmidrule(lr){2-2}
\cmidrule(lr){3-3}
\cmidrule(lr){4-4}
\cmidrule(lr){5-7}
& $-\log(\mathrm{cluster\_p})$ $\uparrow$
& $-\log(\mathrm{cluster\_p})$ $\uparrow$
& $-\log(\mathrm{cluster\_p})$ $\uparrow$
& TPR $\uparrow$
& $-\log(\mathrm{cluster\_p})$ $\uparrow$
& bit\_acc $\uparrow$ \\
\midrule
\multicolumn{7}{l}{\small\textbf{Codec \& Regeneration}} \\
MP3 (64kbps)       & 0.01 & 0.22 & 10.18 & 0.80 & 18.35 & 0.88 \\
Opus (16kbps)      & 0.83 & 2.46 & 22.02 & 0.75 & 25.68 & 0.85 \\
EnCodec (24kbps)    & 1.21 & 2.81 & 23.24 & 0.73 & 26.03 & 0.86 \\
DAC                & 0.78 & 2.57 & 22.83 & 0.70 & 24.62 & 0.80 \\
SpeechTokenizer    & 0.55 & 2.34 & 22.83 & 0.66 & 23.52 & 0.79 \\
FaCodec            & 0.01 & 0.20 &  4.00 & 0.71 & 8.50 & 0.81 \\
VoiceCraft-EnCodec & 0.05 & 0.59 &  9.09 & 0.69 & 13.31 & 0.78 \\
SoCodec            & 0.00 & 0.40 &  3.18 & 0.68 &  6.67 & 0.77 \\
SemantiCodec       & 0.00 & 0.20 &  4.58 & 0.69 &  8.38 & 0.77 \\
\midrule
\multicolumn{7}{l}{\small\textbf{Signal Processing}} \\
high-pass          & 0.02 & 1.42 & 13.06 & 0.90 & 12.32 & 0.93 \\
low-pass           & 0.59 & 1.19 & 22.62 & 0.88 & 25.21 & 0.94 \\
smooth             & 0.08 & 0.54 & 15.74 & 0.89 & 16.98 & 0.92 \\
Gaussian\_noise    & 1.43 & 3.24 & 15.26 & 0.91 & 16.33 & 0.91 \\
pink\_noise        & 1.06 & 3.47 & 25.02 & 0.90 & 23.69 & 0.92 \\
echo               & 1.30 & 3.15 & 25.10 & 0.85 & 25.10 & 0.93 \\
quantization       & 1.38 & 3.41 & 11.06 & 0.83 & 13.57 & 0.92 \\
updownresample     & 1.86 & 3.27 & 25.08 & 0.90 & 25.65 & 0.91 \\
\midrule
\multicolumn{7}{l}{\small\textbf{Desynchronization}} \\
speed              & 0.00 & 0.08 &  5.51 & 0.75 &  9.88 & 0.86 \\
crop               & 1.25 & 2.99 &  0.00 & 0.80 & 11.50 & 0.87 \\
shift (1/8)        & 0.01 & 0.32 &  8.49 & 0.89 & 13.65 & 0.90 \\
shush              & 1.79 & 3.47 & 25.12 & 0.90 & 26.01 & 0.92 \\
\bottomrule
\end{tabular}%
}
\end{subtable}
\end{table*}

\subsection{Ablation Study}

Detailed results in Appendix \ref{ablation}. Codec-union clustering consistently improves post-codec recovery over geometric and retokenization-based alternatives, showing the importance of modeling codec-dependent token substitutions. Multi-lap organization, burst scheduling, and structured decoding each contribute to robustness, while removing the lap pipeline and HiPT search causes the largest performance drop. The complete configuration provides the best overall trade-off between clean extraction and codec robustness.

\section{Conclusion}

We presented MARC, a multi-bit watermarking framework for autoregressive audio generation that integrates codec-aware token clustering, payload-driven scheduling, and alignment-aware verification. Experiments across speech, dialogue, and music generation demonstrate reliable payload recovery under the evaluated codec and overwriting attacks while maintaining audio fidelity, extending token-level watermarking beyond presence detection.

\subsection*{AI use statement}

In this study, we used generative AI tools to refine the text and images in the article.
We did not use generative AI tools to generate images directly.
Furthermore, we used generative AI tools for grammatical correction and to check the article for simple writing errors. We have reviewed all AI-generated outputs. We used AI to perform code syntax checks. We take responsibility for the final content of this study,
including any text, arguments or findings generated using generative AI.





\bibliography{iclr2027_conference}
\bibliographystyle{iclr2027_conference}

\appendix
\section{Appendix}

\subsection{Ablation Study}
\label{ablation}

\paragraph{Key Verification Ablation.}

We evaluate different conditions of the pub-key and pri-key in Table~\ref{tab:public-verification}. Condition A measures acceptance and recovery with both parameters correct, which is the legal verifier.

We examine the complementary roles of the pub-key and pri-key
by evaluating all four combinations of correct and incorrect
keys in Table~\ref{tab:public-verification}. This design separates candidate assessment from public verification: the pri-key scores candidate payloads before aggregation, whereas the pub-key guides post-voting refinement and verifies the resulting payload.

With both keys correct, MARC achieves a TPR@FPR=1\% of 0.94,
a -log($cluster\_p$) of 37.08, and a bit\_acc of 0.98.
Keeping the pub-key correct but replacing the pri-key reduces
these values to 0.58, 8.06, and 0.66, respectively.
Thus, access to the correct pub-key alone does not reproduce
the recovery performance obtained with the correct pri-key,
supporting private-key-guided candidate assessment.
Conversely, using the correct pri-key with an incorrect pub-key
yields a bit\_acc of 0.78 and a TPR@FPR=1\% of 0.76,
demonstrating the additional contribution of public-key-guided
refinement and verification.
When both keys are incorrect, bit\_acc falls to 0.52.
Together, these results support the use of both keys at
complementary decoding stages: private-key scoring identifies
promising candidates, while the pub-key provides an additional
constraint on the aggregated result.

\begin{table*}[h]
\centering
\caption{Identity Verification.}
\label{tab:public-verification}
\begin{tabular}{cllccc}
\toprule
Condition & pub-key & pri-key & TPR@FPR=1\% $\uparrow$ & -log($cluster\_p$)$\uparrow$
& bit\_acc $\uparrow$ \\
\midrule
A & Correct & Correct & 0.94 & 37.08 & 0.98   \\
B & Correct & Incorrect & 0.58 &  8.06&0.66\\
C & Incorrect & Correct & 0.76 &  15.32 & 0.78\\
D & Incorrect & Incorrect &   0.49& 5.08& 0.52\\
\bottomrule
\end{tabular}
\par\smallskip
\begin{minipage}{\textwidth}
\end{minipage}
\end{table*}

\paragraph{Clustering Ablation.}
Table 5 compares the full codec-union construction (C0) with GMM + retokenization (C1), geometric GMM (C2), and GMM + retokenization + CNBP (C3). The cluster count and construction data budget are held fixed. 

Codec transformations can replace a token with another token
that is not its nearest geometric neighbor. MARC therefore
combines intrinsic token geometry with substitution information
from multiple codecs, aiming to preserve encoded symbols across
different reconstruction processes.
Table~\ref{tab:financeqa-ablation-clustering} compares the full
codec-union construction (C0) with GMM + retokenization (C1),
Geometric GMM (C2), and GMM + retokenization + CNBP (C3),
while keeping the cluster count and construction data budget fixed.

On clean audio, C0 achieves the highest bit\_acc of 98.17\%
and the lowest FAD of 0.2622 among the evaluated variants,
showing that incorporating codec information can support recovery
without requiring greater distributional distortion.
Across the eight codec attacks, C0 also achieves the highest
mean bit\_acc of 97.77\%, compared with 97.24\% for C1,
97.41\% for C2, and 97.06\% for C3.
For example, under EnCodec, C0 reaches 99.25\% bit\_acc,
compared with 97.92\%, 98.25\%, and 96.92\%, respectively.

These results support the value of combining substitution
evidence across codecs rather than relying solely on geometric
proximity or the evaluated retokenization-based refinements.
The resulting clusters provide a representation in which
codec-induced token changes can preserve the embedded symbol,
supporting robust payload recovery across heterogeneous codecs.

\begin{table}[htbp]
\centering
\caption{Clustering ablations on FinanceQA.}
\label{tab:financeqa-ablation-clustering}
\begingroup
\small
\setlength{\tabcolsep}{3pt}

\textbf{(a) Clean performance}\par\smallskip
\resizebox{\linewidth}{!}{%
\begin{tabular}{@{}llcccccc@{}}
\toprule
\multirow[c]{2}{*}{ID}
& \multirow[c]{2}{*}{Clustering}
& \multirow[c]{2}{*}{TPR@FPR=1\% $\uparrow$}
& \multirow[c]{2}{*}{$-\log(\mathrm{cluster\_p})$$\uparrow$}
& \multirow[c]{2}{*}{bit\_acc(\%) $\uparrow$}
& \multirow[c]{2}{*}{FAD $\downarrow$}
& \multicolumn{2}{c}{MOS $\uparrow$} \\
\cmidrule(lr){7-8}
& & & & & & NISQA & DNSMOS \\
\midrule
C0 & Full (codec union)          & 94.00 & 37.08 & 99.17 & 0.2622 & 4.503 & 3.868 \\
C1 & GMM + retokenization        & 96.00 & 32.84 & 99.00 & 0.3791 & 4.508 & 3.889 \\
C2 & Geometric GMM              & 95.00 & 36.96 & 98.75 & 0.3799 & 4.512 & 3.800 \\
C3 & GMM + retokenization + CNBP & 91.00 & 35.20 & 98.00 & 0.4137 & 4.462 & 3.855 \\
\bottomrule
\end{tabular}%
}

\medskip
\textbf{(b) Codec robustness}\par\smallskip
\resizebox{\linewidth}{!}{%
\begin{tabular}{@{}ll*{8}{c}@{}}
\toprule
ID & Metric
& \shortstack{SpeechTokenizer\\ \quad}
& \shortstack{EnCodec\\(24 kbps)}
& \shortstack{MP3\\(32 kbps)}
& \shortstack{Opus\\(16 kbps)}
& \shortstack{DAC\\(44.1 kHz)}
& \shortstack{FACodec\\(16 kHz)}
& \shortstack{SoCodec\\ \quad}
& \shortstack{SemantiCodec\\ \quad} \\
\midrule
\multirow{3}{*}{C0}
& TPR@FPR=1\% $\uparrow$      & 90.00 & 97.00 & 87.00 & 96.00 & 94.00 & 92.00 & 79.00 & 85.00 \\
& $-\log(\mathrm{cluster\_p})$$\uparrow$        & 25.33 & 30.70 & 24.11 & 29.20 & 28.84 & 30.29 & 23.03 & 24.02 \\
& bit\_acc(\%) $\uparrow$ & 97.58 & 99.25 & 97.17 & 95.00 & 98.33 & 98.25 & 95.50 & 97.08 \\
\midrule
\multirow{3}{*}{C1}
& TPR@FPR=1\% $\uparrow$      & 86.00 & 92.00 & 82.00 & 92.00 & 95.00 & 94.00 & 70.00 & 89.00 \\
& $-\log(\mathrm{cluster\_p})$$\uparrow$        & 23.67 & 26.98 & 23.99 & 28.12 & 26.65 & 29.35 & 22.99 & 24.02 \\
& bit\_acc(\%) $\uparrow$ & 97.17 & 97.92 & 96.33 & 97.92 & 98.92 & 98.75 & 93.83 & 97.08 \\
\midrule
\multirow{3}{*}{C2}
& TPR@FPR=1\% $\uparrow$      & 88.00 & 92.00 & 86.00 & 97.00 & 95.00 & 92.00 & 76.00 & 80.00 \\
& $-\log(\mathrm{cluster\_p})$$\uparrow$        & 24.56 & 27.95 & 24.33 & 29.43 & 28.50 & 32.26 & 24.17 & 24.18 \\
& bit\_acc(\%) $\uparrow$ & 97.25 & 98.25 & 96.83 & 99.33 & 99.08 & 98.08 & 94.83 & 95.67 \\
\midrule
\multirow{3}{*}{C3}
& TPR@FPR=1\% $\uparrow$      & 85.00 & 87.00 & 82.00 & 91.00 & 92.00 & 91.00 & 77.00 & 86.00 \\
& $-\log(\mathrm{cluster\_p})$$\uparrow$        & 24.90 & 26.10 & 25.06 & 28.53 & 28.89 & 30.09 & 23.40 & 24.33 \\
& bit\_acc(\%) $\uparrow$ & 97.00 & 96.92 & 96.17 & 98.00 & 98.33 & 98.08 & 95.08 & 96.92 \\
\bottomrule
\end{tabular}%
}
\endgroup
\end{table}

\paragraph{Encoding Ablation.}
We compare the full symbol schedule with alternative scheduling and redundancy configurations in table \ref{tab:financeqa-ablation-embedding}. MARC organizes payload embedding into laps and bursts to provide repeated, structured observations for decoding. We evaluate this design in
Table~\ref{tab:financeqa-ablation-embedding} by removing codec
union (E1), using a single lap (E2), setting the burst length
to one (E3), or replacing the full schedule with round-robin
scheduling (E4).

The full configuration achieves a mean post-codec bit\_acc
of 97.77\%, compared with 97.24\% without codec union,
94.71\% with a single lap, 96.42\% with unit bursts,
and 92.41\% with round-robin scheduling.
The full schedule outperforms E2-E4 in both bit\_acc and
TPR@FPR=1\% under every evaluated codec attack.
The benefit of multiple laps is particularly clear under
SemantiCodec, where bit\_acc increases from 91.08\% for E2
to 97.08\% for the full configuration.
Structured scheduling also improves EnCodec recovery from
92.58\% with round-robin scheduling to 99.25\%.

The unit-burst comparison further distinguishes clean extraction
from robustness after reconstruction.
Although E3 achieves 99.42\% bit\_acc on clean audio,
the full configuration provides higher recovery accuracy under
all eight codec attacks, including 99.25\% versus 96.50\%
under EnCodec.
This pattern supports the use of bursts to provide repeated
local evidence, together with multiple laps that distribute
payload observations across the sequence.
The full configuration also has the lowest clean-audio FAD
among the evaluated encoding variants, demonstrating a favorable
combination of recovery robustness and audio fidelity.

\begin{table}[htbp]
\centering
\caption{Embedding ablations on FinanceQA.}
\label{tab:financeqa-ablation-embedding}
\begingroup
\small
\setlength{\tabcolsep}{3pt}

\textbf{(a) Clean performance}\par\smallskip
\resizebox{\linewidth}{!}{%
\begin{tabular}{@{}llcccccc@{}}
\toprule
\multirow[c]{2}{*}{ID}
& \multirow[c]{2}{*}{Embedding}
& \multirow[c]{2}{*}{TPR@FPR=1\% $\uparrow$}
& \multirow[c]{2}{*}{$-\log(\mathrm{cluster\_p})$$\uparrow$}
& \multirow[c]{2}{*}{bit\_acc(\%) $\uparrow$}
& \multirow[c]{2}{*}{FAD $\downarrow$}
& \multicolumn{2}{c}{MOS $\uparrow$} \\
\cmidrule(lr){7-8}
& & & & & & NISQA & DNSMOS \\
\midrule
E0 & Full                   & 94.00 & 37.08 & 99.17 & 0.2622 & 4.503 & 3.868 \\
E1 & w/o codec union        & 96.00 & 32.84 & 99.00 & 0.2791 & 4.508 & 3.889 \\
E2 & Single lap ($L=1$)     & 90.00 & 34.86 & 97.58 & 0.2777 & 4.498 & 3.868 \\
E3 & Unit burst ($B=1$)     & 98.00 & 30.32 & 99.42 & 0.2718 & 4.523 & 3.873 \\
E4 & Round-robin scheduling & 81.00 & 26.35 & 95.92 & 0.2726 & 4.480 & 3.812 \\
\bottomrule
\end{tabular}%
}

\medskip
\textbf{(b) Codec robustness}\par\smallskip
\resizebox{\linewidth}{!}{%
\begin{tabular}{@{}ll*{8}{c}@{}}
\toprule
ID & Metric
& \shortstack{SpeechTokenizer\\ \quad}
& \shortstack{EnCodec\\(24 kbps)}
& \shortstack{MP3\\(32 kbps)}
& \shortstack{Opus\\(16 kbps)}
& \shortstack{DAC\\(44.1 kHz)}
& \shortstack{FACodec\\(16 kHz)}
& \shortstack{SoCodec\\ \quad}
& \shortstack{SemantiCodec\\ \quad} \\
\midrule
\multirow{3}{*}{E0}
& TPR@FPR=1\% $\uparrow$      & 90.00 & 97.00 & 87.00 & 96.00 & 94.00 & 92.00 & 79.00 & 85.00 \\
& $-\log(\mathrm{cluster\_p})$$\uparrow$        & 25.33 & 30.79 & 24.11 & 29.20 & 28.84 & 30.29 & 23.03 & 24.38 \\
& bit\_acc(\%) $\uparrow$ & 97.58 & 99.25 & 97.17 & 99.00 & 98.33 & 98.25 & 95.50 & 97.08 \\
\midrule
\multirow{3}{*}{E1}
& TPR@FPR=1\% $\uparrow$      & 86.00 & 92.00 & 82.00 & 92.00 & 95.00 & 94.00 & 70.00 & 89.00 \\
& $-\log(\mathrm{cluster\_p})$$\uparrow$        & 23.67 & 26.98 & 23.99 & 28.12 & 26.65 & 29.35 & 22.99 & 24.02 \\
& bit\_acc(\%) $\uparrow$ & 97.17 & 97.92 & 96.33 & 97.92 & 98.92 & 98.75 & 93.83 & 97.08 \\
\midrule
\multirow{3}{*}{E2}
& TPR@FPR=1\% $\uparrow$      & 80.00 & 91.00 & 77.00 & 88.00 & 91.00 & 87.00 & 64.00 & 68.00 \\
& $-\log(\mathrm{cluster\_p})$$\uparrow$        & 26.23 & 27.58 & 24.48 & 29.07 & 29.89 & 30.25 & 24.49 & 24.25 \\
& bit\_acc(\%) $\uparrow$ & 94.42 & 97.33 & 94.08 & 97.08 & 97.42 & 96.42 & 89.83 & 91.08 \\
\midrule
\multirow{3}{*}{E3}
& TPR@FPR=1\% $\uparrow$      & 83.00 & 83.00 & 83.00 & 93.00 & 92.00 & 91.00 & 68.00 & 78.00 \\
& $-\log(\mathrm{cluster\_p})$$\uparrow$        & 23.52 & 23.90 & 22.96 & 23.97 & 24.52 & 24.29 & 21.46 & 22.86 \\
& bit\_acc(\%) $\uparrow$ & 96.50 & 96.50 & 96.42 & 98.08 & 97.67 & 98.17 & 92.67 & 95.33 \\
\midrule
\multirow{3}{*}{E4}
& TPR@FPR=1\% $\uparrow$      & 63.00 & 65.00 & 57.00 & 68.00 & 68.00 & 74.00 & 73.00 & 57.00 \\
& $-\log(\mathrm{cluster\_p})$$\uparrow$        & 20.52 & 20.62 & 19.54 & 21.03 & 20.55 & 21.07 & 20.01 & 19.81 \\
& bit\_acc(\%) $\uparrow$ & 92.33 & 92.58 & 90.58 & 93.00 & 92.58 & 94.25 & 93.42 & 90.58 \\
\bottomrule
\end{tabular}%
}
\endgroup
\end{table}

\paragraph{Decoding Ablation.}
Table \ref{tab:financeqa-ablation-decoder} compares the full decoder (D0) with two variants: joint removal of the lap pipeline and HiPT grid search (D1), and removal of boundary sliding (D2).  

Even when watermark evidence survives codec reconstruction,
payload recovery requires assigning that evidence to the correct
symbol positions.
MARC addresses this requirement through structured alignment
search and local boundary correction.

Removing the lap pipeline and HiPT grid search reduces clean
bit\_acc from 99.17\% to 84.75\% and mean post-codec bit\_acc
from 97.77\% to 74.57\%.
The difference is particularly pronounced under SoCodec,
where recovery increases from 64.67\% with D1 to 95.50\%
with the full decoder.
Boundary sliding provides an additional, separately measured
benefit: removing it lowers mean post-codec bit\_acc to 94.84\%,
with SemantiCodec recovery decreasing from 97.08\% to 91.92\%.
The full decoder improves both bit\_acc and TPR@FPR=1\%
over D1 and D2 under every evaluated codec attack.

Notably, -log($cluster\_p$) remains unchanged across the decoder
variants on clean audio and seven of the eight codec attacks,
despite substantial differences in payload recovery.
This distinction demonstrates why strong cluster-level evidence
alone is insufficient for accurate multi-bit decoding.
MARC's decoding pipeline converts preserved watermark evidence
into recoverable payload information by resolving its alignment,
while boundary sliding further corrects local positional errors.

\begin{table}[htbp]
\centering
\caption{Decoder ablations on FinanceQA.}
\label{tab:financeqa-ablation-decoder}
\begingroup
\small
\setlength{\tabcolsep}{3pt}

\textbf{(a) Clean performance}\par\smallskip
\resizebox{\linewidth}{!}{%
\begin{tabular}{@{}llcccccc@{}}
\toprule
\multirow[c]{2}{*}{ID}
& \multirow[c]{2}{*}{Decoder}
& \multirow[c]{2}{*}{TPR@FPR=1\% $\uparrow$}
& \multirow[c]{2}{*}{$-\log(\mathrm{cluster\_p})$$\uparrow$}
& \multirow[c]{2}{*}{bit\_acc(\%) $\uparrow$}
& \multirow[c]{2}{*}{FAD $\downarrow$}
& \multicolumn{2}{c}{MOS $\uparrow$} \\
\cmidrule(lr){7-8}
& & & & & & NISQA & DNSMOS \\
\midrule
D0 & Full
& 94.00 & 37.08 & 99.17 & 0.2622 & 4.503 & 3.868 \\
D1 & w/o lap pipeline and HiPT grid
& 66.00 & 37.08 & 84.75 & 0.3021 & 4.498 & 3.842\\
D2 & w/o boundary sliding
& 92.00 & 37.08 & 97.75 & 0.2932 & 4.486 & 3.874 \\
\bottomrule
\end{tabular}%
}

\medskip
\textbf{(b) Codec robustness}\par\smallskip
\resizebox{\linewidth}{!}{%
\begin{tabular}{@{}ll*{8}{c}@{}}
\toprule
ID & Metric
& \shortstack{SpeechTokenizer\\ \quad}
& \shortstack{EnCodec\\(24 kbps)}
& \shortstack{MP3\\(32 kbps)}
& \shortstack{Opus\\(16 kbps)}
& \shortstack{DAC\\(44.1 kHz)}
& \shortstack{FACodec\\(16 kHz)}
& \shortstack{SoCodec\\ \quad}
& \shortstack{SemantiCodec\\ \quad} \\
\midrule
\multirow{3}{*}{D0}
& TPR@FPR=1\% $\uparrow$      & 90.00 & 97.00 & 87.00 & 96.00 & 94.00 & 92.00 & 78.00 & 85.00 \\
& $-\log(\mathrm{cluster\_p})$$\uparrow$        & 25.33 & 30.79 & 24.11 & 29.20 & 28.84 & 30.29 & 23.03 & 24.02 \\
& bit\_acc(\%) $\uparrow$ & 97.58 & 99.25 & 97.17 & 99.00 & 98.33 & 98.25 & 95.50 & 97.08 \\
\midrule
\multirow{3}{*}{D1}
& TPR@FPR=1\% $\uparrow$      & 38.00 & 53.00 & 31.00 & 52.00 & 56.00 & 54.00 & 19.00 & 26.00 \\
& $-\log(\mathrm{cluster\_p})$$\uparrow$        & 25.33 & 30.79 & 24.11 & 29.20 & 28.84 & 30.29 & 23.03 & 23.84 \\
& bit\_acc(\%) $\uparrow$ & 74.25 & 80.67 & 71.00 & 79.83 & 81.33 & 79.08 & 64.67 & 65.75 \\
\midrule
\multirow{3}{*}{D2}
& TPR@FPR=1\% $\uparrow$      & 81.00 & 84.00 & 74.00 & 86.00 & 85.00 & 81.00 & 66.00 & 65.00 \\
& $-\log(\mathrm{cluster\_p})$$\uparrow$        & 25.33 & 30.79 & 24.11 & 29.20 & 28.84 & 30.29 & 23.03 & 24.05 \\
& bit\_acc(\%) $\uparrow$ & 95.42 & 96.25 & 94.17 & 96.83 & 96.25 & 95.67 & 92.25 & 91.92 \\
\bottomrule
\end{tabular}%
}

\smallskip
\endgroup
\end{table}

\subsection{Supplementary Robustness Experiments}
\label{Robustness Experiments}

We have included here some experimental data on other parameter settings not covered in the main table. As shown in Table \ref{tab:robustness-extra-financeQA}, Table \ref{tab:robustness-extra-lib} and Table \ref{tab:robustness-extra-dolly}.

For conventional audio attacks. We first evaluate the robustness of MARC during audio transmission and processing by means of conventional audio attacks. Such attacks perturb the audio through operations such as noise addition, filtering, compression and temporal editing, with the aim of weakening the watermark evidence or disrupting the correspondence between embedding and extraction. This experiment examines whether the watermark can still be detected after the audio content has undergone common processing, and whether the carried message can be correctly recovered, thereby providing a basis for assessing the practical usability of the method.

In conventional attacks, we focus on the codec attacks. Codec attacks introduce structural token substitutions and alignment changes by re-representing and synthesizing audio, thereby directly challenging watermarking methods that rely on discrete speech representations. Even if the reconstructed audio retains a similar auditory quality, the re-extracted token sequence may undergo significant changes. 

MARC addresses this issue by organizing watermark information within a symbolic representation at the cluster level, enabling partial token changes to retain the same watermark meaning, whilst compensating for local evidence loss through repeated observation and aggregated decoding. This experiment aims to verify whether these designs can transform instability at the token level into stability at the message level, thereby supporting detection and recovery following codec transformation. 

We further examine attribution interference when an attacker actively introduces another watermark message. We therefore consider an overwriting attack, the aim of which is to weaken the original owner’s watermark evidence whilst simultaneously establishing valid support for the attacker’s message, thereby causing the audio to be erroneously attributed to the attacker. Consequently, this experiment must simultaneously examine the preservation of the original watermark and the effectiveness of the attacker’s watermark insertion: the failure of the original watermark indicates removal, the acceptance of both watermarks indicates coexistence, whilst the failure of the original watermark coupled with the validity of the attacker’s watermark constitutes a successful replacement. 

For MARC, this evaluation focuses on whether multi-bit message recovery and corresponding verification can distinguish between different attribution claims, and to what extent the distributed evidence of the original message can be preserved under active interference. Combined with codec attacks, the two types of experiments respectively test information retention capability under representational transformations and attribution reliability under competing watermarks, reflecting the method’s design objectives for audio re-encoding and active watermark replacement.

\begin{table*}[thb]
\centering
\captionsetup{skip=2pt}
\caption{Robustness Supplementary.}
\label{tab:robustness-extra-financeQA}
\setlength{\tabcolsep}{3pt}
\renewcommand{\arraystretch}{1}
\begin{subtable}{\textwidth}
\centering
\caption{SpiritLM with FinanceQA}
\resizebox{\linewidth}{!}{%
\begin{tabular}{l*{6}{c}}
\toprule
\multirow[c]{2}{*}{Attacks}
& AR speech WM & Aligned-IS & HiPT & \multicolumn{3}{c}{MARC} \\
\cmidrule(lr){2-2}
\cmidrule(lr){3-3}
\cmidrule(lr){4-4}
\cmidrule(lr){5-7}
& $-\log_{10}(\mathrm{cluster\_p})$ $\uparrow$
& $-\log_{10}(\mathrm{cluster\_p})$ $\uparrow$
& $-\log_{10}(\mathrm{cluster\_p})$ $\uparrow$
& TPR@FPR=1\% $\uparrow$
& $-\log_{10}(\mathrm{cluster\_p})$ $\uparrow$
& bit\_acc $\uparrow$ \\
\midrule
\multicolumn{7}{l}{\small\textbf{Codec \& Regeneration}} \\
\addlinespace[2pt]
MP3 (32 kbps) & 2.148 & 3.436 & 13.310 & 87.00 & 24.107 & 97.17 \\
MP3 (40 kbps) & 2.193 & 3.111 & 13.065 & 86.00 & 24.217 & 96.83 \\
Opus (31 kbps) & 3.330 & 4.390 & 15.140 & 97.00 & 31.112 & 99.25 \\
Opus (32 kbps) & 3.538 & 4.374 & 14.537 & 92.00 & 30.836 & 98.25 \\
Opus (64 kbps) & 3.524 & 4.474 & 14.621 & 94.00 & 31.254 & 98.92 \\
EnCodec (3k) & 2.254 & 3.724 & 13.262 & 91.00 & 25.269 & 97.83 \\
EnCodec (6k) & 3.050 & 4.243 & 13.255 & 92.00 & 25.653 & 97.92 \\
EnCodec (12k) & 3.201 & 4.490 & 13.766 & 95.00 & 28.162 & 98.83 \\
\midrule
\multicolumn{7}{l}{\small\textbf{Signal Processing}} \\
\addlinespace[2pt]
Gaussian noise ($\sigma$=0.01) & 2.702 & 3.901 & 14.322 & 85.00 & 21.319 & 96.75 \\
\midrule
\multicolumn{7}{l}{\small\textbf{Desynchronization}} \\
\addlinespace[2pt]
Shift (1/2) & 1.091 & 2.619 & 12.968 & 75.00 & 21.915 & 94.58 \\
\bottomrule
\end{tabular}%
}
\end{subtable}

\vspace{2pt}
\end{table*}

\begin{table*}[thb]
\centering
\captionsetup{skip=2pt}
\caption{Robustness Supplementary.}
\label{tab:robustness-extra-lib}
\setlength{\tabcolsep}{3pt}
\renewcommand{\arraystretch}{1}
\begin{subtable}{\textwidth}
\centering
\caption{SpiritLM with LibriSpeech}
\resizebox{\linewidth}{!}{%
\begin{tabular}{l*{6}{c}}
\toprule
\multirow[c]{2}{*}{Attacks}
& AR speech WM & Aligned-IS & HiPT & \multicolumn{3}{c}{MARC} \\
\cmidrule(lr){2-2}
\cmidrule(lr){3-3}
\cmidrule(lr){4-4}
\cmidrule(lr){5-7}
& $-\log_{10}(\mathrm{cluster\_p})$ $\uparrow$
& $-\log_{10}(\mathrm{cluster\_p})$ $\uparrow$
& $-\log_{10}(\mathrm{cluster\_p})$ $\uparrow$
& TPR@FPR=1\% $\uparrow$
& $-\log_{10}(\mathrm{cluster\_p})$ $\uparrow$
& bit\_acc $\uparrow$ \\
\midrule
\multicolumn{7}{l}{\small\textbf{Codec \& Regeneration}} \\
\addlinespace[2pt]
MP3 (32 kbps) & 1.531 & 3.605 & 8.905 & 87.00 & 21.939 & 97.08 \\
MP3 (40 kbps) & 1.463 & 3.568 & 8.881 & 90.00 & 21.989 & 97.75 \\
Opus (31 kbps) & 2.982 & 5.454 & 9.401 & 92.00 & 25.477 & 97.92 \\
Opus (32 kbps) & 3.092 & 5.256 & 9.697 & 91.00 & 26.375 & 97.83 \\
Opus (64 kbps) & 3.032 & 5.115 & 9.645 & 93.00 & 26.592 & 98.25 \\
EnCodec (3k) & 1.751 & 4.725 & 9.860 & 91.00 & 22.415 & 97.83 \\
EnCodec (6k) & 2.599 & 4.944 & 10.070 & 85.00 & 23.156 & 96.42 \\
EnCodec (12k) & 2.876 & 5.121 & 9.755 & 86.00 & 23.336 & 97.00 \\

\midrule
\multicolumn{7}{l}{\small\textbf{Signal Processing}} \\
\addlinespace[2pt]
Gaussian noise ($\sigma$=0.01) & 2.649 & 4.431 & 10.003 & 86.00 & 19.792 & 96.75 \\
\midrule
\multicolumn{7}{l}{\small\textbf{Desynchronization}} \\
\addlinespace[2pt]
Shift (1/2) & 0.809 & 2.751 & 9.296 & 77.00 & 20.744 & 95.00 \\
\bottomrule
\end{tabular}%
}
\end{subtable}

\vspace{2pt}
\end{table*}
\begin{table*}[thb]
\centering
\captionsetup{skip=2pt}
\caption{Robustness Supplementary.}
\label{tab:robustness-extra-dolly}
\setlength{\tabcolsep}{3pt}
\renewcommand{\arraystretch}{1}
\begin{subtable}{\textwidth}
\centering
\caption{SpiritLM with Dolly}
\resizebox{\linewidth}{!}{%
\begin{tabular}{l*{6}{c}}
\toprule
\multirow[c]{2}{*}{Attacks}
& AR speech WM & Aligned-IS & HiPT & \multicolumn{3}{c}{MARC} \\
\cmidrule(lr){2-2}
\cmidrule(lr){3-3}
\cmidrule(lr){4-4}
\cmidrule(lr){5-7}
& $-\log_{10}(\mathrm{cluster\_p})$ $\uparrow$
& $-\log_{10}(\mathrm{cluster\_p})$ $\uparrow$
& $-\log_{10}(\mathrm{cluster\_p})$ $\uparrow$
& TPR@FPR=1\% $\uparrow$
& $-\log_{10}(\mathrm{cluster\_p})$ $\uparrow$
& bit\_acc $\uparrow$ \\
\midrule
\multicolumn{7}{l}{\small\textbf{Codec \& Regeneration}} \\
\addlinespace[2pt]
MP3 (32 kbps) & 1.596 & 3.945 & 17.639 & 84.00 & 23.721 & 96.00 \\
MP3 (40 kbps) & 1.820 & 3.861 & 17.025 & 87.00 & 24.118 & 97.00 \\
Opus (31 kbps) & 2.988 & 5.324 & 17.639 & 91.00 & 29.413 & 98.00 \\
Opus (32 kbps) & 3.094 & 5.393 & 17.639 & 95.00 & 29.557 & 98.75 \\
Opus (64 kbps) & 3.214 & 5.459 & 17.807 & 95.00 & 29.250 & 98.83 \\
EnCodec (3k) & 1.975 & 4.250 & 17.451 & 87.00 & 24.138 & 97.58 \\
EnCodec (6k) & 2.486 & 4.918 & 17.947 & 90.00 & 25.051 & 98.17 \\
EnCodec (12k) & 2.980 & 5.123 & 17.874 & 93.00 & 25.448 & 98.50 \\
\midrule
\multicolumn{7}{l}{\small\textbf{Signal Processing}} \\
\addlinespace[2pt]
Gaussian noise ($\sigma$=0.01) & 2.363 & 5.150 & 16.887 & 85.00 & 20.805 & 96.58 \\
\midrule
\multicolumn{7}{l}{\small\textbf{Desynchronization}} \\
\addlinespace[2pt]
Shift (1/2) & 0.866 & 2.710 & 17.085 & 78.00 & 22.579 & 95.17 \\
\bottomrule
\end{tabular}%
}
\end{subtable}

\vspace{2pt}
\end{table*}

\paragraph{Supplementary robustness evaluation.}
We extend the main evaluation with additional compression
bitrates, additive noise, and temporal shifts on FinanceQA,
LibriSpeech, and Dolly.
Tables~\ref{tab:robustness-extra-financeQA},
\ref{tab:robustness-extra-lib}, and
\ref{tab:robustness-extra-dolly} report the results.
After excluding duplicate codec configurations, MARC achieves
the highest $-\log_{10}(\mathrm{cluster\_p})$ among all compared
methods, attack settings, while maintaining
bit\_acc between 94.58\% and 99.25\%.
These results extend the main findings across both parameter
settings and datasets, demonstrating the preservation of
watermark evidence alongside accurate multi-bit recovery.

\paragraph{Robustness across compression settings.}
Compression can change the token sequence even when the
reconstructed audio preserves its perceptual content.
The additional MP3, Opus, and EnCodec settings therefore
test whether MARC's recovery depends on a particular codec
or bitrate.
Across all codec configurations evaluated over the
three datasets, MARC maintains bit\_acc of 96.00-99.25\%
and TPR@FPR=1\% of 84.00-97.00\%.
Under MP3 at 32~kbps, bit\_acc remains at 97.17\% on
FinanceQA, 97.08\% on LibriSpeech, and 96.00\% on Dolly.
For Opus at 31, 32, and 64~kbps, all nine dataset bit\_acc of at least 97.83\%,
showing consistently accurate recovery across the tested
operating points.

The neural-codec results provide further evidence for
MARC's representation-level design.
Even at an EnCodec bandwidth of 3~kbps, MARC achieves
bit\_acc of 97.83\%, 97.83\%, and 97.58\% on FinanceQA,
LibriSpeech, and Dolly, respectively.
Its corresponding $-\log_{10}(\mathrm{cluster\_p})$ scores
are 25.269, 22.415, and 24.138, compared with
13.262, 9.860, and 17.451 for HiPT, the strongest
baseline on this metric in these settings.
These results are consistent with the motivation for
multi-codec clustering: substitutions within a cluster
can preserve the encoded symbol despite changes in
individual token identities.
Repeated observations and aggregated decoding further
support recovery when some symbol evidence is disrupted.

\paragraph{Robustness to noise and temporal offsets.}
The additional Gaussian-noise condition uses a fixed
standard deviation of 0.01, testing recovery under an
absolute perturbation scale.
MARC retains bit\_acc of 96.75\% on FinanceQA,
96.75\% on LibriSpeech, and 96.58\% on Dolly,
while achieving the highest
$-\log_{10}(\mathrm{cluster\_p})$ on all three datasets.
This demonstrates that the observed robustness extends
beyond codec reconstruction to direct waveform corruption.

The half-frame shift tests a different requirement:
assigning preserved evidence to the correct payload
positions after temporal displacement.
Under this condition, MARC achieves bit\_acc of
94.58\%, 95.00\%, and 95.17\% on the three datasets.
Its cluster-evidence scores also exceed those of all
baselines in each case.
Accurate recovery under this offset is consistent with
the role of alignment-aware decoding and boundary
correction in resolving positional mismatch.

\paragraph{Consistency across datasets.}
The same pattern holds across FinanceQA, LibriSpeech,
and Dolly: MARC combines stronger cluster-level evidence
than the evaluated baselines with high payload recovery
accuracy under every supplementary setting.
This consistency supports the complementary design of
cluster-level encoding, repeated payload scheduling,
and alignment-aware decoding.
Together, they address changes in token identity,
partial corruption of symbol evidence, and displacement
of extraction positions, extending watermark robustness
from presence detection to recoverable multi-bit information.

\subsection{Attack Settings}
\label{Settings}

Let $\mathbf{x}\in\mathbb{R}^{T}$ denote a watermarked waveform
sampled at $f_s=16$~kHz.
Each attack is applied independently to the original waveform:
\begin{equation}
    \widetilde{\mathbf{x}}
    =\mathcal{A}_{\theta}(\mathbf{x}),
\end{equation}
where $\theta$ specifies the attack parameters.
We evaluate both watermark evidence and payload recovery,
since retaining detectable evidence does not necessarily
preserve its correspondence to message positions.

\paragraph{Signal Processing.}
These attacks model acoustic interference, bandwidth restriction,
and waveform processing during transmission and redistribution.
They test whether watermark information survives changes
in acoustic content and the resulting token assignments.

\begin{itemize}
    \item \textbf{Gaussian noise.}
    We evaluate two conditions: an SNR of $30$~dB and a
    fixed noise standard deviation of $0.01$.
    The attacked waveform is
    \begin{equation}
        \widetilde{x}_t=x_t+n_t,
        \qquad n_t\sim\mathcal{N}(0,\sigma^2).
    \end{equation}
    For the SNR-controlled condition,
    \begin{equation}
        \sigma^2=
        \left(\frac{1}{T}\sum_{t=0}^{T-1}x_t^2\right)
        10^{-30/10}.
    \end{equation}
    These conditions test recovery under broadband
    interference at relative and absolute noise scales.

    \item \textbf{Pink noise.}
    We use the implementation's pink-noise routine,
    which constructs a correlated sequence $\mathbf{z}$
    from cumulative sums of Gaussian samples.
    The sequence is peak-normalized before addition:
    \begin{equation}
        \widetilde{\mathbf{x}}
        =
        \mathbf{x}
        +0.01\frac{\mathbf{z}}{\|\mathbf{z}\|_{\infty}}.
    \end{equation}
    Here, $0.01$ specifies the peak noise amplitude
    rather than its standard deviation.
    This condition complements independent Gaussian noise
    by introducing temporally correlated interference.

    \item \textbf{High-pass and low-pass filtering.}
    We apply Julius high-pass and low-pass filters with
    cutoff frequencies of $1$~kHz and $3$~kHz,
    respectively.
    Their normalized cutoffs are $1000/16000$ and
    $3000/16000$.
    These attacks model frequency-selective processing
    and test recovery after different acoustic components
    are suppressed.

    \item \textbf{Smoothing.}
    We apply a six-sample moving-average filter:
    \begin{equation}
        z_t=\frac{1}{6}\sum_{j=0}^{5}x_{t+j},
        \qquad 0\leq t<T-5.
    \end{equation}
    The valid filtering output is right-padded with
    zeros to restore length $T$.
    This models local waveform smoothing and tests
    sensitivity to the removal of rapid acoustic variations.

    \item \textbf{Echo.}
    We apply a two-tap kernel with coefficients $1$
    and $0.05$, separated by $800$ samples ($50$~ms),
    using \texttt{julius.fft\_conv1d}.
    The filtered output is rescaled to the original
    peak amplitude and right-padded to length $T$.
    This condition tests recovery under delayed
    acoustic interference while controlling the
    output peak level.

    \item \textbf{Quantization.}
    We uniformly quantize waveform amplitudes to
    $64$ levels between the waveform minimum $a$
    and maximum $b$:
    \begin{equation}
        \widetilde{x}_t
        =
        a+\frac{b-a}{63}
        \operatorname{round}
        \left(63\frac{x_t-a}{b-a}\right).
    \end{equation}
    Constant waveforms are left unchanged.
    This models reduced amplitude precision during
    storage or processing.

    \item \textbf{Up/down resampling.}
    We resample from $16$ to $32$~kHz and then back
    to $16$~kHz using Julius:
    \begin{equation}
        \widetilde{\mathbf{x}}
        =
        \mathcal{R}_{32\rightarrow16}
        \left(
        \mathcal{R}_{16\rightarrow32}(\mathbf{x})
        \right).
    \end{equation}
    The output retains the original sample count.
    This tests the effects of interpolation and filtering
    during sample-rate conversion between audio systems.
\end{itemize}

MARC addresses these perturbations through cluster-level
encoding: a token substitution need not change the encoded
symbol when cluster membership is preserved.
Repeated payload observations and cross-lap aggregation
provide additional recovery evidence when individual
observations are corrupted.
This category therefore assesses recovery without requiring
exact preservation of the original waveform.

\paragraph{Desynchronization.}
These attacks model temporal editing and partial observation.
They challenge either the correspondence between embedding
and extraction positions or the availability of evidence
at those positions.

\begin{itemize}
    \item \textbf{Speed perturbation.}
    We use \texttt{librosa.effects.time\_stretch}
    with rates $\alpha\in\{0.9,1.1\}$.
    This operation changes duration while approximately
    preserving pitch.
    The transformed waveform is then matched to the
    original length:
    \begin{equation}
        \widetilde{\mathbf{x}}
        =
        \mathcal{M}_{T}
        \left(
        \operatorname{TS}_{\alpha}(\mathbf{x})
        \right).
    \end{equation}
    The $1.1$ condition adds trailing zero padding,
    whereas the $0.9$ condition truncates the stretched
    signal.
    These settings model playback-rate editing and test
    recovery under temporal distortion and its associated
    boundary effects.

    \item \textbf{Cropping.}
    We retain the first $50\%$ of the waveform without
    padding:
    \begin{equation}
        \widetilde{\mathbf{x}}
        =\mathbf{x}[0:\lfloor T/2\rfloor].
    \end{equation}
    This models excerpt sharing and partial audio
    availability, testing whether the retained segment
    contains sufficient evidence for payload recovery.

    \item \textbf{Temporal shifts.}
    We circularly shift the waveform to the right by
    $1/8$ or $1/2$ of a $320$-sample frame:
    \begin{equation}
        \widetilde{x}_t=x_{(t-\Delta)\bmod T},
        \qquad \Delta\in\{40,160\}.
    \end{equation}
    These offsets correspond to $2.5$ and $10$~ms.
    Circular shifting preserves all samples while
    changing their alignment with extraction boundaries,
    isolating sensitivity to temporal offsets.

    \item \textbf{Shush.}
    We replace the first $1\%$ of samples with zeros:
    \begin{equation}
        \widetilde{x}_t=
        \begin{cases}
            0, & t<\lfloor0.01T\rfloor,\\
            x_t, & \text{otherwise}.
        \end{cases}
    \end{equation}
    This models onset muting and tests resilience to
    missing initial evidence.
    Unlike cropping, it preserves the waveform length
    and the positions of subsequent samples.
\end{itemize}

MARC distributes repeated payload observations across laps
to support recovery from incomplete evidence.
Alignment search and boundary sliding address positional
mismatch, while cross-lap aggregation combines surviving
observations.
These experiments test whether watermark evidence remains
usable for message recovery after temporal editing.

\paragraph{Codec \& Regeneration.}
These attacks model lossy storage, bandwidth-constrained
transmission, and reconstruction through alternative audio
representations.
For codec $c$ with configuration $r$, the round trip is
\begin{equation}
    \widetilde{\mathbf{x}}
    =
    \operatorname{Dec}_{c,r}
    \left(
    \operatorname{Enc}_{c,r}(\mathbf{x})
    \right).
\end{equation}

\begin{itemize}
    \item \textbf{MP3.}
    We evaluate encoding and decoding at $32$, $40$,
    and $64$~kbps.
    These settings test robustness across compression
    strengths encountered during storage and redistribution.

    \item \textbf{Opus.}
    We evaluate round-trip compression at $16$, $31$,
    $32$, and $64$~kbps.
    These configurations model different bandwidth
    constraints in communication and audio delivery.

    \item \textbf{EnCodec.}
    We evaluate round-trip reconstruction at $3$, $6$,
    $12$, and $24$~kbps using the $24$~kHz codec.
    These settings test recovery after neural encoding,
    quantization, and reconstruction at different
    supported bandwidths.

    \item \textbf{DAC.}
    We use the default $44.1$~kHz DAC model.
    The input is resampled from $16$ to $44.1$~kHz
    before reconstruction, and the output is returned
    to $16$~kHz and matched to the original length.
    This evaluates an alternative neural-codec
    representation without imposing an additional
    bitrate constraint.

    \item \textbf{SpeechTokenizer.}
    We perform tokenization followed by waveform
    reconstruction.
    This tests whether payload information survives
    processing through a different speech-token
    representation.

    \item \textbf{FaCodec.}
    We apply the FaCodec reconstruction backend.
    This adds another learned analysis, synthesis
    transformation to the evaluation.

    \item \textbf{VoiceCraft-EnCodec.}
    We perform an encoding-decoding round trip using
    VoiceCraft's four-codebook EnCodec checkpoint.
    This evaluates its codec component rather than
    text-conditioned speech editing or generation.

    \item \textbf{SoCodec.}
    We apply SoCodec analysis, synthesis at $16$~kHz.
    The wrapper prepends $8000$ zero samples
    ($0.5$~s) before processing and removes the
    corresponding prefix from the reconstructed output.
    The remaining waveform is matched to the
    original length.

    \item \textbf{SemantiCodec.}
    We use the configuration with a token rate of $50$
    and a semantic vocabulary size of $16384$.
    The waveform is encoded and decoded using the
    corresponding checkpoint, and the output is
    matched to the original length.
\end{itemize}

Codec reconstruction can preserve perceptual content while
changing the token sequence used for watermark extraction.
MARC addresses this challenge by combining intrinsic token
geometry with multi-codec substitution information.
For an original token $y_t$ and its reconstructed counterpart
$\widetilde{y}_{\pi(t)}$, cluster-level encoding permits
\begin{equation}
    \widetilde{y}_{\pi(t)}\neq y_t
    \quad\text{while}\quad
    c(\widetilde{y}_{\pi(t)})=c(y_t),
\end{equation}
where $\pi(t)$ denotes the temporal correspondence when
one exists.
Within-cluster substitutions therefore preserve the encoded
symbol.
Together with repeated embedding and alignment-aware decoding,
this design targets payload recovery across heterogeneous
reconstruction processes.

\end{document}